 \documentstyle[epsf,twocolumn,aps]{revtex}

\begin{document}

\topmargin = -0.0cm

\draft

\title{Constructing transportable behavioural models for nonlinear electronic
devices} 

\author{David M. Walker}
\address{Department of Applied Science,
College of William and Mary, Williamsburg VA 23187-8795 \\
HP Labs, MS4-AD, 1501 Page Mill Rd, Palo Alto CA 94304-1126}

\author{Reggie Brown}
\address{Department of Physics and Department of Applied Science,
College of William and Mary, Williamsburg VA 23187-8795}

\author{Nicholas B. Tufillaro}
\address{HP Labs, MS4-AD, 1501 Page Mill Rd, Palo Alto CA 94304-1126 \\
Department of Applied Science, College of William and Mary, Williamsburg 
VA 23187-8795}

\date{\today}

\maketitle

\begin{abstract}
We use radial basis functions to model the input--output response of
an electronic device.  A new methodology for producing models that
accuratly describe the response of the device over a wide range of
operating points is introduced.  A key to the success of the method
is the ability to find a polynomial relationship between the model
parameters and the operating points of the device.
\end{abstract}

\pacs{05.45.+b}


\section{Introduction}
In this letter we investigate the possibility of applying methods
recently developed for modeling nonlinear chaotic systems to modelling
input--output dynamics of a nonlinear electronic device.
The ability to describe the behaviour of modern electronic 
devices is becoming increasingly important.  As technology advances,
and the demands on current technology increase, it is becoming more 
and more apparent that traditional linear modelling methods fail to 
describe the behaviour of modern electronic devices~\cite{jiang}.
This is particularly true when the device is being operated outside 
of its ``normal'' (linear) response regime, and is subject to 
complicated (nonlinear) driving signals~\cite{hasler,banerjee}.

It has been emphasized by researchers in nonlinear dynamics that
complicated behaviour, which appears random, can sometimes arise
from a low dimensional deterministic nonlinear
system~\cite{hdia,casdagli,hdia2}.  It seems apt, then, to investigate
whether nonlinear modelling techniques can supplement traditional 
linear modelling techniques when they fail to describe a devices 
behaviour~\cite{mjc,book,bk}.

One result of our study is a new methodology for modelling electronic 
devices.  In particular, we focus on constructing {\em transportable} 
behavioural models of such devices.  In this research the dynamics of
the device is represented by input--output data obtained at a few
operating regimes.  A behavioural model is constructed by adjusting 
its parameters until it accurately mimics the dynamics of the device.
Within this context, a behavioural model is transportable if it
accurately mimics the dynamics of the device in operating regimes
which are different from the ones used to construct the model.  The
ultimate goal is a single behavioural model that accurately mimics
the device in all of its operating regimes.

Ideally we should compare the performance of our models to that of
``standard'' models using data obtained from a real device.  However,
testing with a real device can be expensive.  Therefore, in the absence 
of real data, we shall regard the ``standard'' model for such a device
as reality.  If our behavioural models match the ``standard'' model
to within a few percent then we claim that we have shown the
{\em potential} of our methods.

\begin{figure}
 \vspace{0.0in}
 \begin{center}
 \leavevmode
 \hbox{%
 \epsfxsize=2.0in 
 \epsfysize=2.0in
 \epsffile{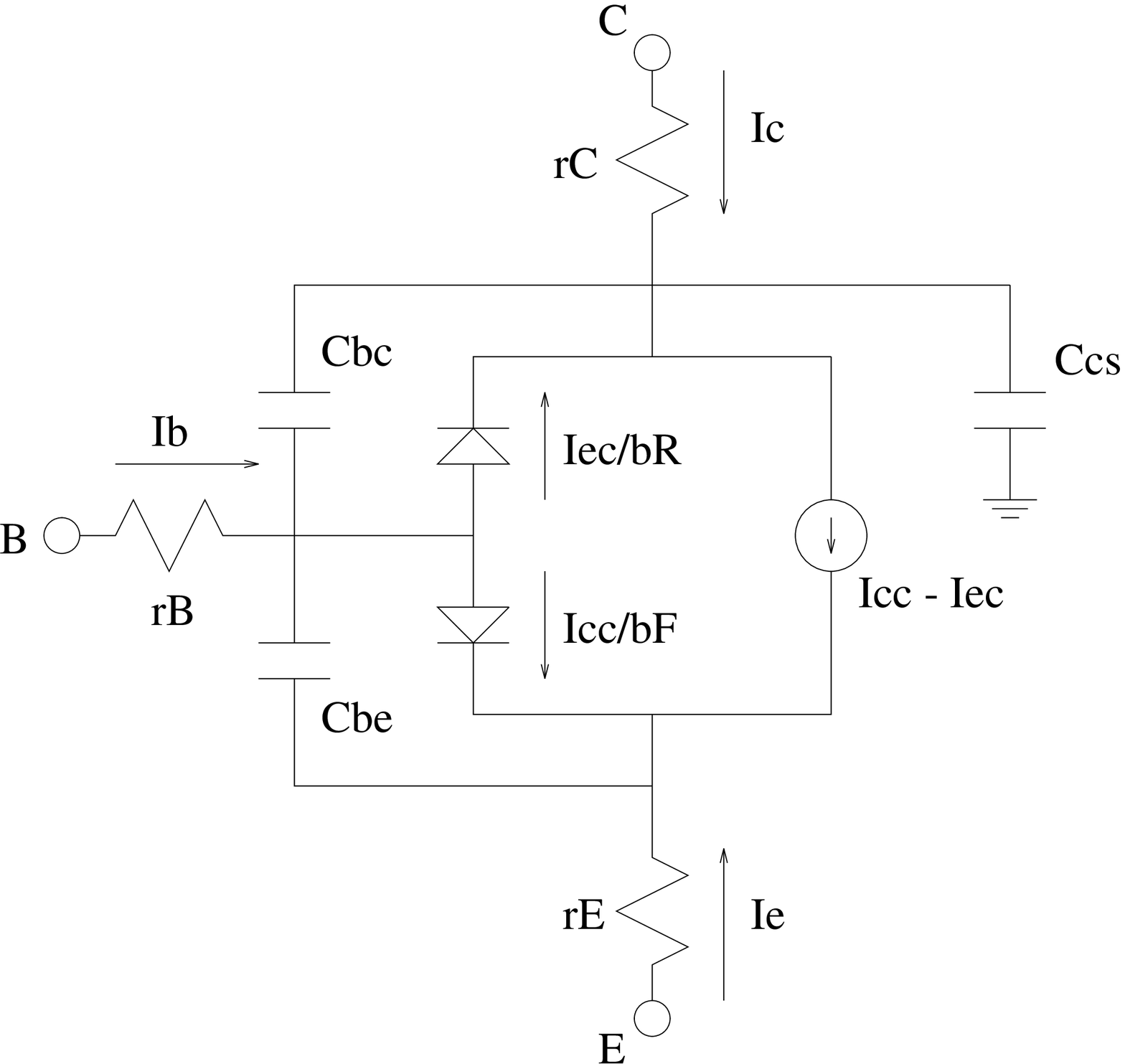}}
 \end{center}
 \caption{Modified Ebers-Moll transistor model.
 \label{figure1}}
\end{figure}

\begin{figure}
 \vspace{0.0in}
 \begin{center}
 \leavevmode
 \hbox{%
 \epsfxsize=2.0in
 \epsffile{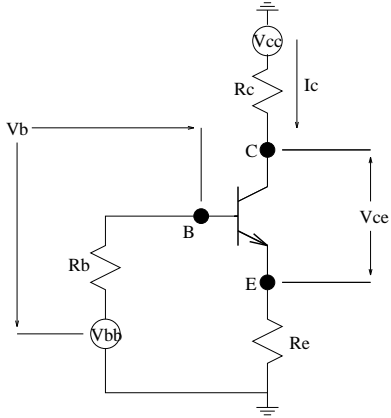}}
 \end{center}
 \caption{Self-biasing transistor circuit.
 \label{figure2}}
\end{figure}

To illustrate the method we will study a device with mild
nonlinearities: a simplification of the ``standard'' Ebers--Moll 
model for a bi-polar junction transistor, which we embed in a 
self-biasing circuit~\cite{self} (see Figs.~\ref{figure2} and 
\ref{figure1}).  That is, we are using the transistor as a simple
amplifier.  We will construct transportable behavioural models
using time domain input--output data obtained from the Ebers--Moll
model~\cite{ebers,mjc}.

\begin{figure}
 \vspace{0.0in}
 \begin{center}
 \leavevmode
 \hbox{%
 \epsfxsize=2.0in 
 \epsffile{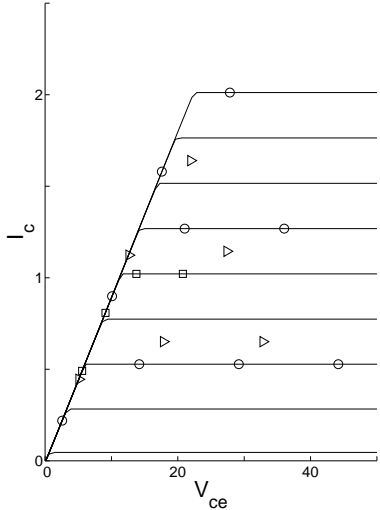}}
 \end{center}
 \caption{dc-characteristics of Ebers-Moll transistor model.  Each
characteristic curve corresponds to $V_{bb}$ ranging from $1$ to $9$
volts.  Points along a particular curve corresponds to $V_{cc}$
ranging from $0$ to $60$ volts.
 \label{figure3}}
\end{figure}

In Fig.~\ref{figure2}, $V_{bb}$ and $V_{cc}$ are potential sources.
If we apply dc-potentials at these points (so called biasing) then
we are selecting an operating point for the device.  A graphical
summary of this part of the devices behaviour is given by the dc 
$V-I$~characteristics obtained by biasing the device at different 
potentials.  Figure~\ref{figure3} shows the $V-I$~characteristics of
our Ebers--Moll transistor model.

The work reported in this letter does not attempt to describe a method 
for constructing completely transportable behavioural models.  A
complete study of transportability would include an investigation of
the models ability to describe a device operating at {\em all bias 
points} and subjected to {\em all possible classes} of drive signals.
Because, the drive signal a device ``sees'' is very much application
dependent, and in lieu of knowledge of a specific application, we will
investigate a restricted notion of transportability.  We demonstrate
our methods for the device operating at many different dc-bias potentials,
bu subjected to the same class of amplitude modulated (AM) drive signals.
Thus, in this letter we investigate transportability across cd-bias 
potentials and avoid the issue of transportability across signal class.

\begin{figure}
 \vspace{0.0in}
 \begin{center}
 \leavevmode
 \hbox{%
 \epsfxsize=2.0in
 \epsffile{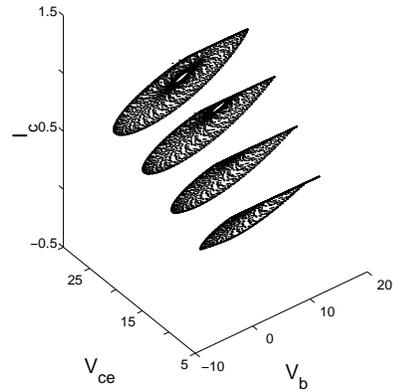}}
 \end{center}
 \caption{Four input-output data sets (pancakes) along the
characteristic curve corresponding to $V_{bb}=5V$, indicated by the
 squares in Fig.~\protect{\ref{figure3}}.
 \label{figure4}}
\end{figure}

The class of AM drive signal we investigate is~\cite{self}
\begin{equation}
\label{eq:am}
V = [(1 + A \sin (\omega t)) V_{c}] \; \sin (\Omega t) ,
\end{equation}
with $V_{c}=5V$, $\omega=5$MHz, $\Omega=500$MHz, and $A=4/5$.  This 
signal is added to the fixed dc-offset at $V_{bb}$ (see Fig.~\ref{figure2})
with fixed dc-offset at $V_{cc}$ (see Fig.~\ref{figure2}).
The input--output data we work with is obtained by integrating the 
circuit equations for the Ebers--Moll model of Fig.~\ref{figure1}
embedded in the amplifier shown in Fig.~\ref{figure2}.
Approximately $N=3000$ input--output data points are generated from
these equations by integrating them from $0$ to $3 \mu s$, sampling 
every $1 ns$.  The inputs for the behavioural models are potential 
differences, $(V_{b},\: V_{ce})$, the output data is the current, 
$I_{c}$.  In Fig.~\ref{figure4} we show examples of input--output
data generated at the four different offset bias points indicated
by the squares in Fig.~\ref{figure3}.  (We will refer to these data 
sets as pancakes.) 


\section{Nonlinear Modelling}
The nonlinear behavioural models we reconstruct are radial basis
function models.  They attempt to predict the measured current, 
$I_{c}$ (the output) from the measured potential differences, $V_{b}$
and $V_{ce}$ (the inputs).  In passing, we remark that the inputs to
the behavioural models are not the same as the AM signal used to drive
the device.

The models have the following form
\begin{equation}
\label{eq1}
I_{c} (t) = F \left[ V_{b} (t), V_{ce} (t) \right],
\end{equation}
where $F\left[ u(t) \right]$ is a radial basis function with affine
terms,
\begin{equation}
\label{def_F}
F\left[ u(t) \right] = \beta + \alpha \: u(t) + \sum_{i=1}^{K}
\omega_{i} \: \phi( \| c_{i} - u(t) \|).
\end{equation}
Here, $u(t)=\left[ V_{b}(t), V_{ce}(t) \right]$, while $\beta$, 
$\alpha$, and $\omega_{i}$ are parameters to be estimated.  The 
$c_{i}$'s are called centres and also need to be estimated.  The
function $\phi$ is a Gaussian
\begin{displaymath}
\phi(r) =  \exp \left( -\frac{r^2}{2\sigma^2} \right) ,
\end{displaymath}
where $\sigma$ is another parameter which is kept fixed at $\sigma
= 4$ (corresponding to half the standard deviation of the $V_{ce}$
data) throughout our investigations.

For many cases it may be necessary to embed the voltages, $V_{b}$
and $V_{ce}$, using their past as well as present values in the 
function approximation of Eq.~(\ref{eq1}).  This will certainly be 
the case for complicated drive signals.  However, for the class of
AM driving signals we examined, the device produces signals which
are simple enough that embedding can be avoided. 

When using radial basis functions an important issue is how many
centres should be used, and where should they be located.  An 
effective method of choosing appropriate centres from a large
set of candidates is the subset selection method of Judd and
Mees~\cite{jees}.  This method attempts to find the {\em best} 
selection of centres from a set of candidates.  It selects the 
best set by evaluating a description length criteria.  We will 
use this method (subject to the Schwarz Information Criterion) to 
determine (in some sense) an optimal model structure using data
collected at one bias point.


\subsection{Constructing Simple Models}
\begin{figure}
 \vspace{0.0in}
 \begin{center}
 \leavevmode
 \hbox{%
 \epsfxsize=2.0in
 \epsffile{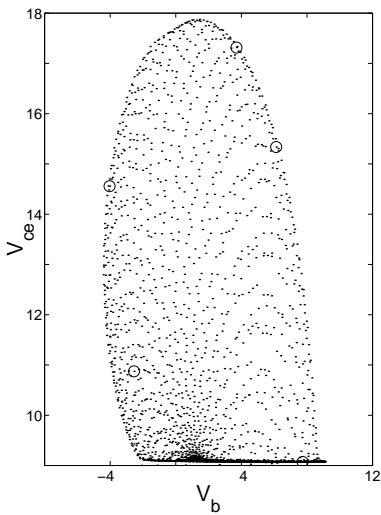}}
 \end{center}
 \caption{The input space of the data generated at $(V_{bb},V_{cc})
 =(5,18)$V.  The five circles are the locations of data points chosen
 as centres by the modelling method.
 \label{figure5}}
\end{figure}

To illustrate the effectiveness of using radial basis functions to
model the input-output relationship between $(V_{b}$, $V_{ce})$ and
$I_{c}$, consider the data from the bias point at $V_{bb}=5V$ and
$V_{cc}=18V$.  In Fig.~\ref{figure5} we show the input space for this 
data.  (The full input-output space corresponding to this data is the
second pancake from the
bottom of Fig.~\ref{figure4}.)

The modeling method was given every fifth data point as a candidate 
centre.
\begin{figure}
 \vspace{0.0in}
 \begin{center}
 \leavevmode
 \hbox{%
 \epsfxsize=2.0in
 \epsffile{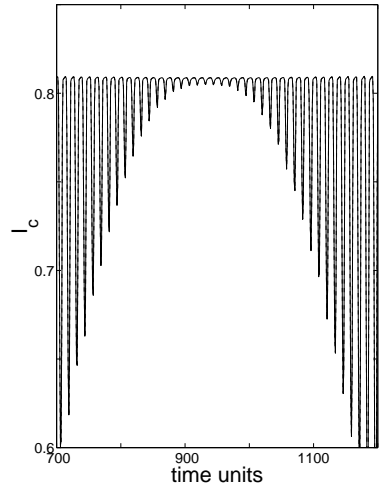}}
 \end{center}
 \caption{A section of the response signal and the predicted signal of
 the nonlinear model reconstructed at $(V_{bb},V_{cc})=(5,18)V$.  We
 see that the fit is exceptional with the errors too difficult to see.
 \label{figure6}}
\end{figure}

The modeling method kept the linear terms in Eq.~(\ref{def_F}) plus 
five radial basis terms.  The data points selected as centres are 
indicated by the circles in Fig.~\ref{figure5}.  Figure~\ref{figure6} 
shows {\em both} the true $I_c$ output and the output predicted by the
behavioural model.  At this scale the two curves are indistinguishable,
with $RMS = 6.13 \times 10^{-4}$ or $[RMS/{\rm STD}(I_c)] =  0.003$.
Clearly, for this data set a very good fit is possible with a small
number of centres.  We argue that because the modeling method chose
nonlinear terms the data is better described by
a nonlinear model than a purely linear one.


\subsection{Constructing Transportable Models}
Given the effectiveness of our method at constructing nonlinear models
at {\em one} operating point, our goal now is to deliver a single 
nonlinear model which ``works'' over a wide range of operating points.

There are two approaches which can be followed.  The first is to 
develop ``training'' data which covers as wide a range of the devices
input--output space as possible, and then construct a model using 
this data.  The second approach is to generate ``identical'' data 
sets at many bias points, reconstruct models at each bias point, and 
form a global model by interpolating between the reconstructed models.
We have attempted both approaches and have, so far, found the latter
approach to be more fruitful.  Therefore, it will be described below.

In general, the relationship between the modeling parameters and
the bias points is
\begin{eqnarray*}
 \beta & = & \beta \left( V_{bb},V_{cc} \right) \\
 \alpha & = & \alpha \left( V_{bb},V_{cc} \right) \\
 \omega_i & = & \omega_i \left( V_{bb},V_{cc} \right) \\
 c_i & = & c_i \left( V_{bb},V_{cc} \right).
\end{eqnarray*}
We will determine these functional relationships by constructing models
from input--output data corresponding to a few different bias points.
Once this step is completed, parameter values at bias points not 
explicitly in the ``training'' data can be estimated via interpolation
or extrapolation.
In this fashion, we can deliver a nonlinear model that is transportable
over a wide regime of bias points.  The information needed for simulating
the dynamics of a device is the operating bias point (from which we 
determine the parameters of the model) and the input signal sequence.

\begin{figure}
 \vspace{0.0in}
 \begin{center}
 \leavevmode
 \hbox{%
 \epsfxsize=2.0in
 \epsffile{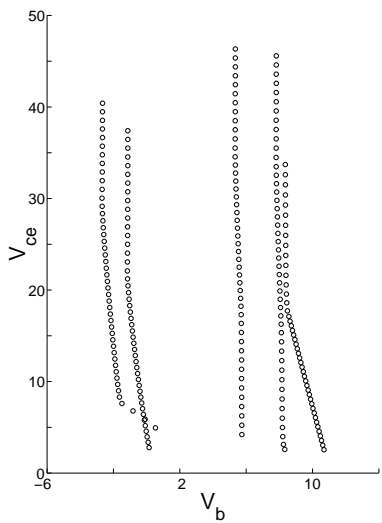}}
 \end{center}
 \caption{The evolution of the five centres chosen by description length
 along the characteristic curve.  We observe that their locations 
 smoothly change with changes in the bias points.
 \label{figure7}}
\end{figure}

Referring once more to Fig.~\ref{figure4} we notice that the pancakes
appear to change smoothly with changes in the operating bias point.  In
Fig.~\ref{figure7} we show that the location of the centres shown in 
Fig.~\ref{figure5} slowly vary as we move along the characteristic curve
corresponding to $V_{bb}=5V$.  We have found that this behaviour is true
for all of the bias curves in Fig.~\ref{figure3}. Also, the data shown
in Fig.~\ref{figure4} was generated in an ``identical'' fashion at each
bias point.  That is, the input--output data point with index~50 (say)
in one pancake corresponds to the data point with index~50 in any other
pancake.  Thus, if the number of centres is fixed, and if the centres in
each pancake have the same indices, then the nonlinear models constructed
with these centres will have the same structure at every bias point.

When constructing a transportable model we keep the number of centres,
and their indices, fixed.  The exact location of a centre is given 
by curves like those in Fig.~\ref{figure7}.  Therefore, only a least 
squares estimate of the weight parameters is necessary.  Implicit in our
approach is the assumption that a smooth change in the bias points will
result in a smooth change in the response signal, which will manifest 
itself as a smooth change in the estimated model parameters.

\begin{figure}
 \vspace{.5in}
 \leavevmode
 \begin{center}
 \epsfxsize=5cm 
 (a) \epsfbox{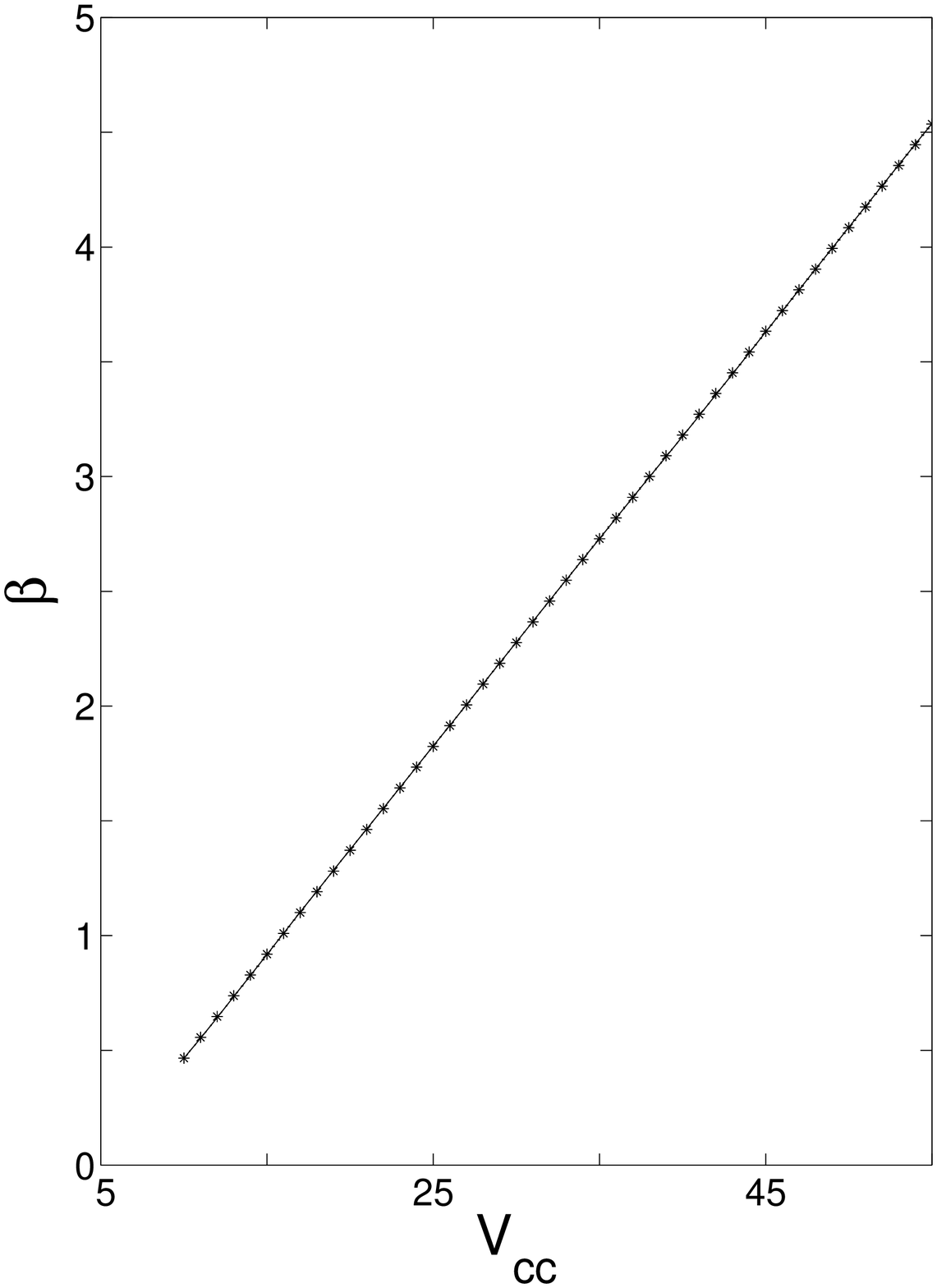}
 \epsfxsize=5cm 
 (b) \epsfbox{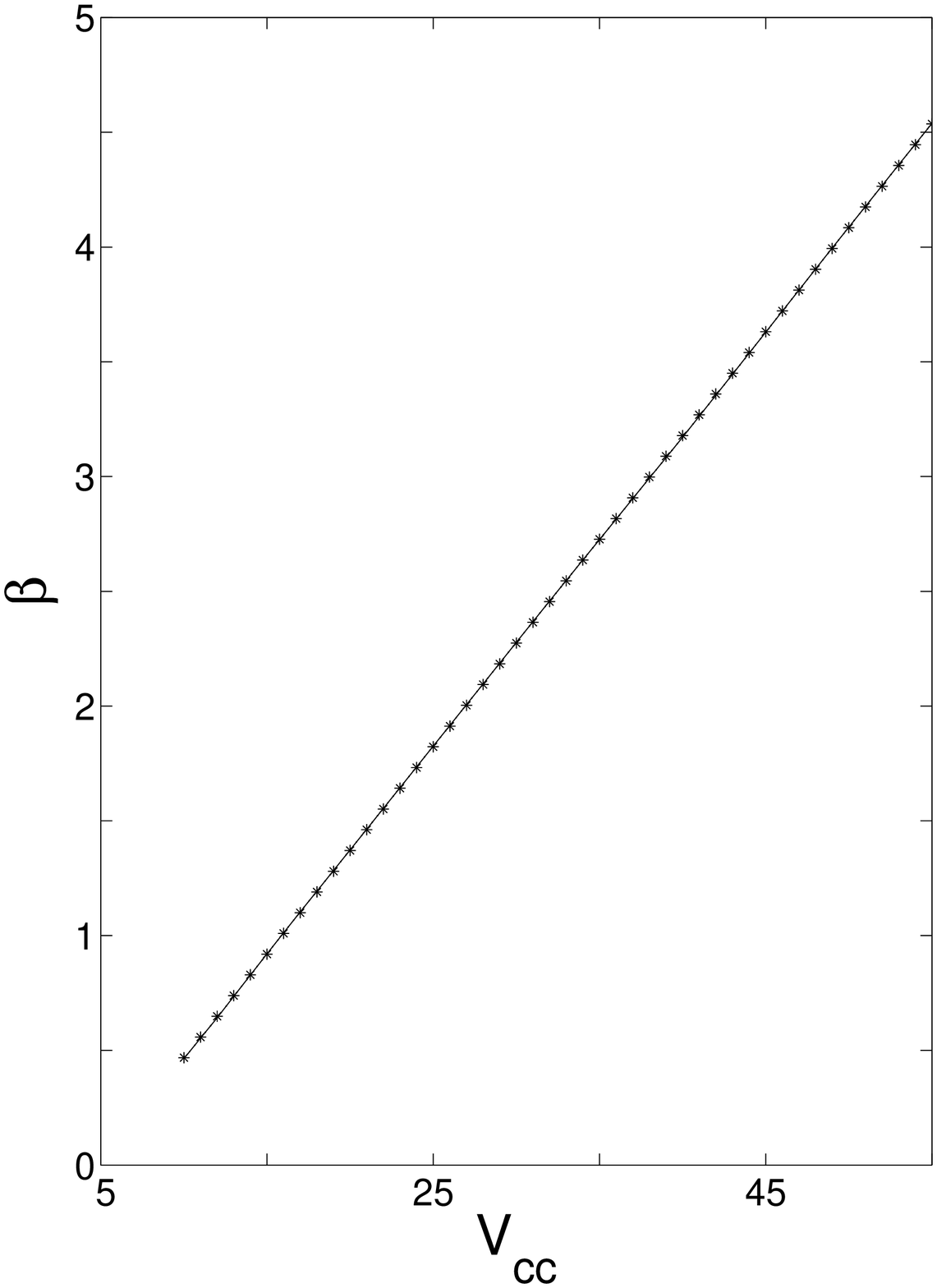} \\
 \epsfxsize=5cm 
 (c) \epsfbox{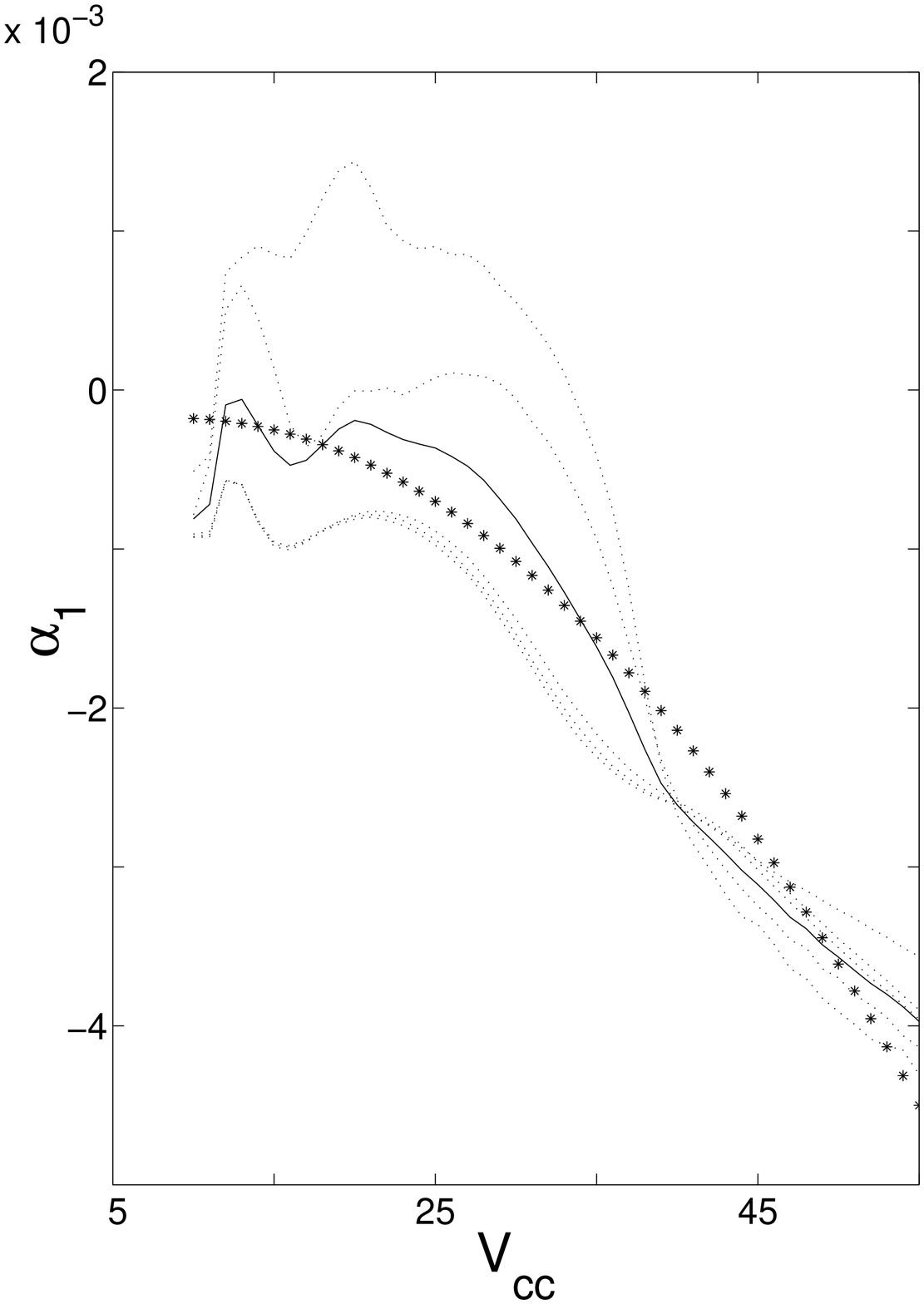}
 \epsfxsize=5cm 
 (d) \epsfbox{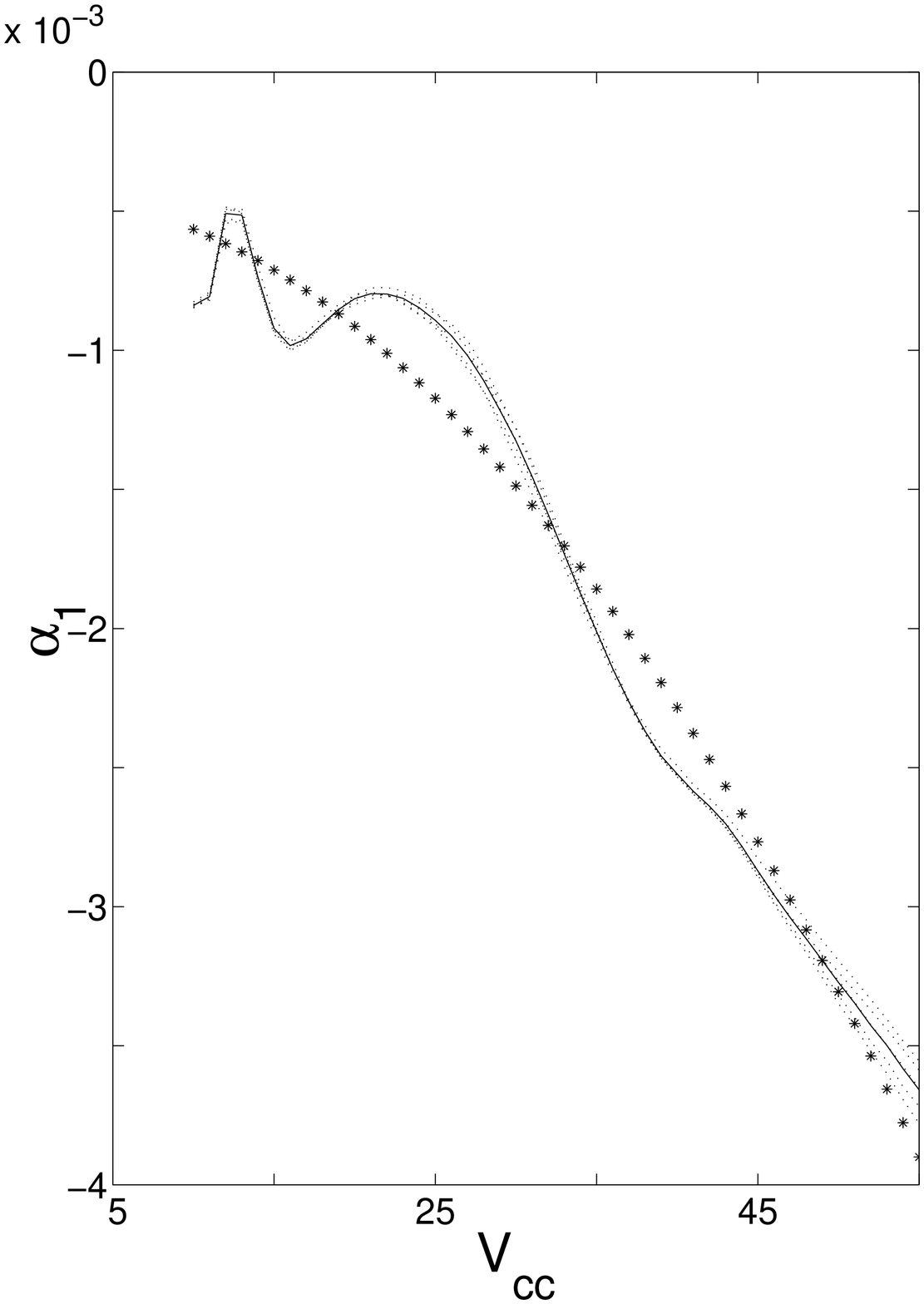} \\
 \epsfxsize=5cm 
 (e) \epsfbox{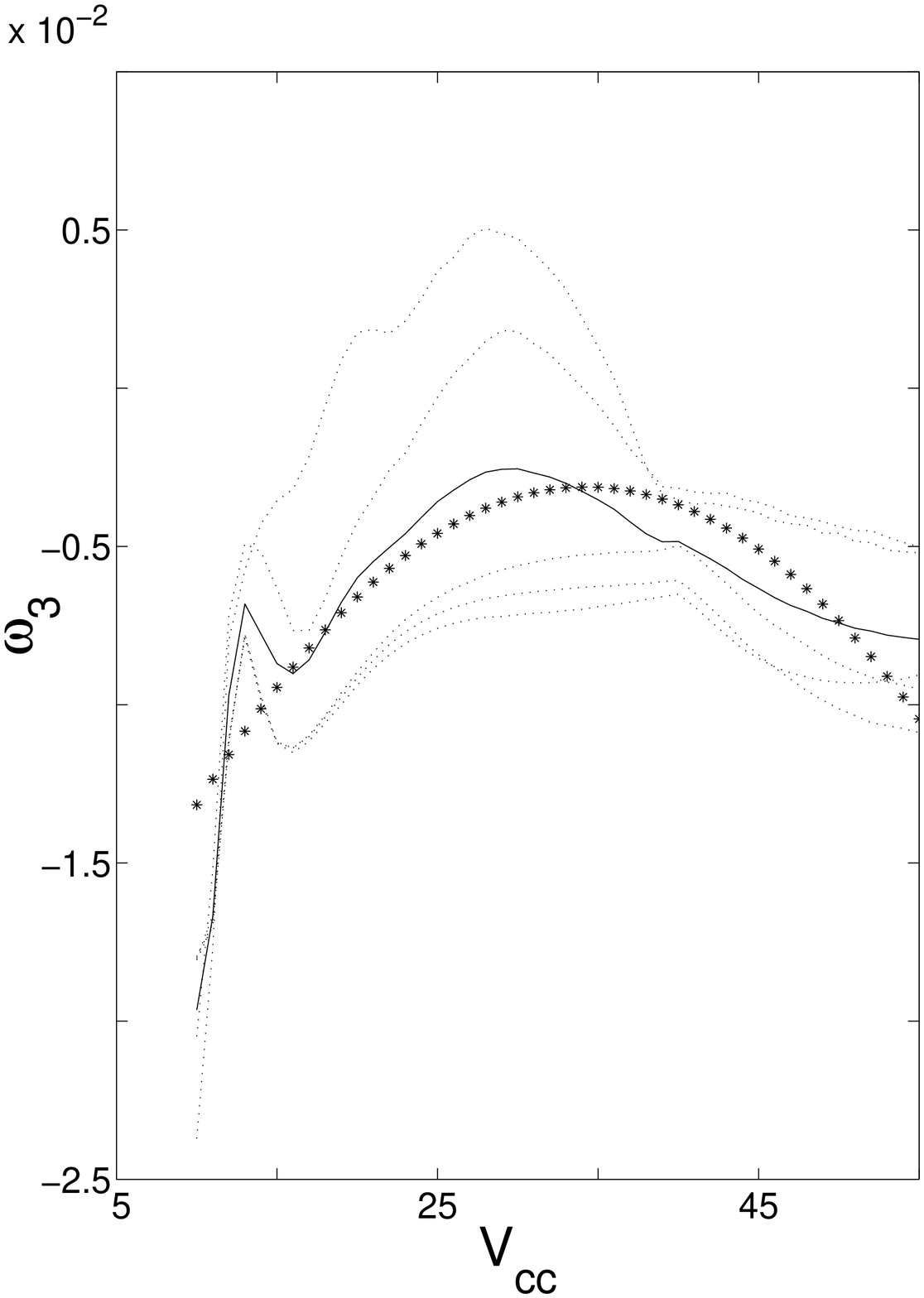}
 \epsfxsize=5cm 
 (f) \epsfbox{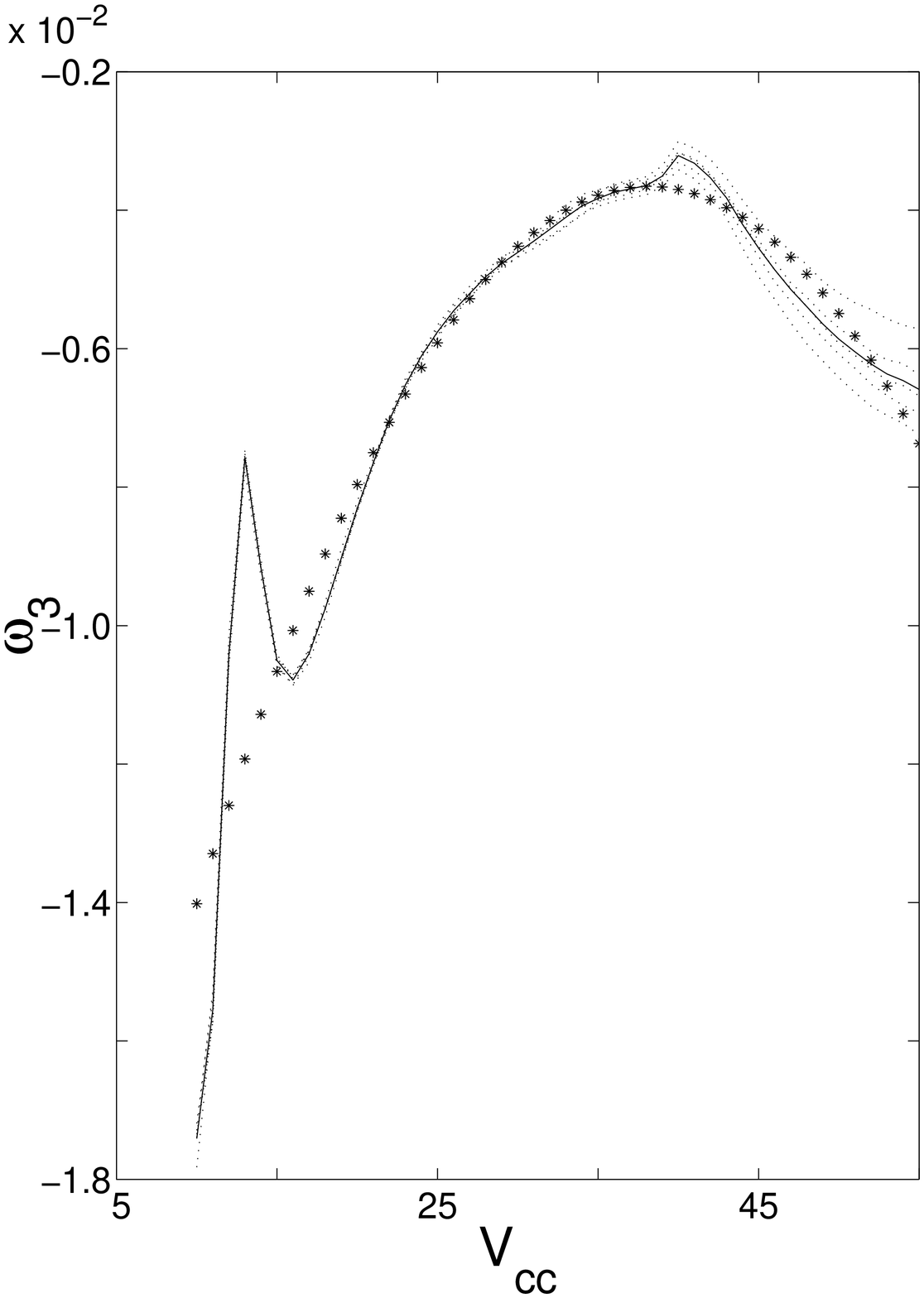} \\
 \end{center}
 \caption{The evolution of the estimated parameters along the
characteristic curve corresponding to $V_{bb}=5V$.  The dotted lines
 show the estimates obtained from the chopped (a), (c) and
 (e), and the decimated (b), (d) and (f) data sets.  The
 solid line is the average of the dotted lines.  Also shown (starred)
line) is the quadratic fit to the parameters using the bias points.}
 \label{figure8}
\end{figure}

This is indeed the case as can be seen in Figs.~\ref{figure8}.
Figures~\ref{figure8}(a) and~\ref{figure8}(b) show two ways of
obtaining an average estimate of the evolution of the constant weight
$\beta$ along the characteristic curve corrsponding to $V_{bb}=5V$
($V_{cc}=5-50V$).  Figures~\ref{figure8}(c) and~\ref{figure8}(d) are
the results for one of the linear weights $\alpha_{1}$, and
Figs.~\ref{figure8}(e) and~\ref{figure8}(f) show the same for a
nonlinear weight, $\omega_{3}$.

To understand how these estimates were obtained, recall that at each
sample bias point we generated a $N=3000$ point data set.  At each
bias point we used two methods to divide this large set into five 
different $600$ point data sets.  In the first method, the first 
$600$ points of the large set forms the first small set, the next 
$600$ points forms the second small set, and so on.  A (five centre)
radial basis mode was constructed for each of the smaller data sets.
The five estimates for the radial basis parameters are given by the 
dotted lines in Figs.~\ref{figure8}(a), (c) and (e).  The solid lines
in the figures are the average values of the parameter estimates, while
the starred lines are quadratic polynomial fits to the average values.

In the second method we decimate the data by taking every $5^{th}$
point to obtain the $600$ point data sets.  The rest of the procedure
for obtaining the model parameters shown in Figs.~\ref{figure8}(b), (d)
and (f) is the same as that just outlined.

The point of the above is to emphasize that the parameter estimates
themselves have ``error bars'' and although the quadratic fits to the
average are bad in places, they nevertheless describe the parameter
evolution quite well as our final results show.


\subsection{New methodology}
Our investigation leads us to propose the following methodology for
producing a transportable model for the electronic device.
\begin{enumerate}
 \item Sample the space of bias voltages, and generate input--output
data sets, in an ``identical'' manner, at each bias point.

 \item Choose one of the data sets to reconstruct a model using the
subset selection and description length methodology proposed by
Judd and Mees~\cite{jees}.

 \item For each input--output data set, reconstruct a nonlinear model
(using centres whose indices were determined in step~2) and store the 
estimated parameters together with their corresponding bias points.

 \item Fit a polynomial relationship between the estimated parameters
and the bias points.
\end{enumerate}
Models constructed using this methodology should be transportable 
over a range of operating points.

To use the behavioural model to predict the input--output dynamics
of the device at one particular bias point, one estimates the values
of the parameters in the model
by interpolating the relationships between the parameters and the bias
point.  Predicted responses to drive signals are generated by inserting
the drive signal into the radial basis model.

We have used the above methodology to produce a transportable radial
basis model of the circuit discussed above.  Behavioural models were
obtained from data generated at twelve bias points ($V_{bb}=3$, 6 and 9,
$V_{cc}=5$, 20, 35, and 50) corresponding to four points on each of
three characteristic curves.  The locations of these points are shown
by circles in Fig.~\ref{figure3}.  Because Fig.~\ref{figure3} is a two
dimensional projection of a three dimensional input--output space there
are only nine circles indicating twelve bias points.  Bias points $(3,5)$, 
$(6,5)$ and $(9,5)$ are actually indicated by the same circle, and 
similary for $(6,20)$ and $(9,20)$.

At each bias point an 
AM driving signal was added as described above.  The indices for the
centres shown in Fig.~\ref{figure5} were used for all of the models,
and a quadratic polynomial was used to describe the relationship
between the model parameters and the bias points.

The transportability of the models was tested using input--output data
generated at nine bias points.  These points are indicated by triangles
in Fig.~\ref{figure3}.  Once again and for the same reasons as above,
bias points $(3.5,10)$, $(5.5,10)$ and $(7.5,10)$ are represented by
the same triangle.  (Similary for $(5.5,25)$ and $(7.5,25)$.)  We
emphasize that this test data represents bias 
points that are completely different from those used to ``train'' the
model.  The results of these out of sample tests are shown in 
Table~\ref{table1}.  We see that good performance is achieved and 
transportability of the model is clearly exhibited over this wide range of 
operating points. 
\begin{table}
\begin{tabular}{|c|c|c|} \hline
 Test set & RMS & RMS/std($I_{c}$) \\
 $(V_{bb},V_{cc})$ & & \\
\hline
 $(3.5,10)$ & $0.0053$ & $0.0324$ \\
\hline
 $(3.5,25)$ & $0.0047$ & $0.0143$ \\
\hline
 $(3.5,40)$ & $0.0099$ & $0.0278$ \\
\hline
 $(5.5,10)$ & $0.0052$ & $0.0412$ \\
\hline
 $(5.5,25)$ & $0.0017$ & $0.0061$ \\
\hline
 $(5.5,40)$ & $0.0129$ & $0.0361$ \\
\hline
 $(7.5,10)$ & $0.0046$ & $0.0850$ \\
\hline
 $(7.5,25)$ & $0.0050$ & $0.0296$ \\
\hline
 $(7.5,40)$ & $0.0070$ & $0.0231$ \\
\hline
\end{tabular}
    \caption{The errors produced by the transportable model for twelve
out of sample test sets.
    \label{table1}}
\end{table}


\section{Conclusion}
We have introduced a new methodology for modelling the input--output
dynamics of nonlinear electronic devices.  Our study was particularly 
focused on producing a nonlinear model which was transportable over a
wide range of operating points.  We used radial basis functions to 
model the device.  Transportability was achieved
by finding a relationship between the models parameters and the
operating points of the device.  A key to the success was the fixing
of model structure (i.e., fixing appropriate centres) so that
interpolation of the parameters with respect to the bias points was
possible.  A topic for future work is to apply our methods to
additional electronic devices.  A second is to investigate the ``other
half'' of the transportability question, namely, can the response of
the device subject to other classes of drive signals be predicted 
accurately by our models.


\section*{Acknowledgment}
This work was supported by the NSF GOALI program under grant number
PHY-9724707 and the AFOSR under grant number F49620-98-1-0144.


\onecolumn

\end{document}